\documentstyle[aps,pra,psfig,amstex]{revtex}
\newcommand{\rar}{\rightarrow}
\newcommand{\al}{\alpha}
\newcommand{\be}{\beta}
\newcommand{\st}{\scriptstyle}
\begin{document}
\preprint{M\'exico ICN-UNAM 99-03}
\title{One-electron linear systems in a strong magnetic field}
\author{ J. C. L\'opez V. and  A. Turbiner\cite{byline}\cite{byline2}}
\address{Instituto de Ciencias Nucleares, UNAM, Apartado Postal 70-543,
        04510 M\'exico D.F., M\'exico}
\date{\today}

\maketitle

\smallskip

\begin{abstract}
  Using a variational method we study a sequence of the one-electron
  atomic and molecular-type systems $H$, $H_2^+$, $H_3^{(2+)}$ and
  $H_4^{(3+)}$ in the presence of a homogeneous magnetic field ranging
  $B = 0 - 4.414\times 10^{13}\,G$. These systems are taken as a linear
  configuration aligned with the magnetic lines. For $H_3^{(2+)}$ the
  potential energy surface has a minimum for $B\sim 10^{11}\,G$ which
  deepens with growth of the magnetic field strength ({\it JETP Lett.
    \bf 69}, 844 (1999)); for $B \gtrsim 10^{12}\,G$ the minimum of the
  potential energy surface becomes sufficiently deep to have
  longitudinal vibrational state.  We demonstrate that for the
  $(ppppe)$ system the potential energy surface at $B \gtrsim
  4.414\times 10^{13}\,G$ develops a minimum, indicating the possible
  existence of exotic molecular ion $H_4^{(3+)}$. We find that for
  almost all accessible magnetic fields $H_2^+$ is the most bound
  one-electron linear system while for magnetic fields $B \gtrsim
  10^{13}\,G$ the molecular ion $H_3^{(2+)}$ becomes the most bound.
\end{abstract}



\widetext

\section{Introduction}
In the presence of a strong magnetic field the properties of atomic
and molecular systems change substantially leading to new unexpected
phenomena. It has been a long time since the predictions made by
Kadomtsev and Kudryavtsev \cite{Kadomtsev:1971}, and Ruderman
\cite{Ruderman:1971}, about the possible existence of unusual chemical
compounds in the presence of a strong magnetic field which are aligned
along the magnetic line. This phenomenon can be explained by the
effective quasi-one-dimensionality of Coulombic systems appearing due
to the enormous Lorentz force acting on a system in transverse
direction.  The transverse size of such systems is governed by the
magnetic field and is of the order of the cyclotron radius ${\hat\rho}
\sim \gamma^{-1/2}$ while the longitudinal size remains of atomic
(molecular) order shrinking logarithmically $\sim 1/\log \gamma$ (see,
for example, \cite{LL})\footnote{Here $\gamma=B/B_0$, where
  $B_0=2.3505\times 10^9\,G$, is in $Ry$ and, in particular, it has a
  meaning of the energy of free electron in magnetic field.}.  As a
consequence of such quasi-one-dimensionality the electron-nuclei
attraction becomes more effective, yielding the possibility of
compensating the Coulombic repulsion of nuclei and thus to the
possible existence of exotic systems that do not exist without a
magnetic field. In particular, it was conjectured that the presence of
a strong magnetic field can lead to the formation of linear molecules
(linear chains) situated along magnetic lines
\cite{Ruderman:1971,Salpeter:1992,Salpeter:1995,Salpeter:1997}.
Recently, the first quantitative study of the possible existence of
the exotic molecular ion $H_3^{(2+)}$ or, in other words, a bound
state in the system of three protons and an electron $(pppe)$ was
carried out \cite{Turbiner:1999}. It turned out that for $B>10^{11}\,G$
the potential energy surface as a function of internuclear distances
possesses a minimum, which deepens with magnetic field growth. It
provides an evidence that for magnetic fields $B > 10^{11}\,G$ such
molecular ion can exist \cite{Turbiner:1999}. In the present article
we extend this study and show that for magnetic fields $B > 10^{13} \,G$
a bound state in $(ppppe)$ system or, in other words, another exotic
molecular ion $H_4^{(3+)}$ can exist.

The aim of this article is to perform a detailed quantitative study
and comparison of the one-electron systems $H$, $H_2^+$, $H_3^{(2+)}$
and $H_4^{(3+)}$ in strong magnetic field. Our study is limited to an
exploration of the ground state only.  It is assumed that the
Born-Oppenheimer approximation holds, which implies that the positions
of protons are kept fixed.  We consider the linear molecular
configuration for which the protons are aligned with the magnetic
field (linear configuration parallel to a magnetic line). This
configuration looks optimal from a physical point of view. Spin
effects (linear Zeeman effect) are neglected. The magnetic field
ranges from 0 up to $4.414\times 10^{13}\, G$, where non-relativistic
considerations hold (for a detailed discussion see, for instance,
\cite{Salpeter:1997} and references therein).  In order to study the
stability of the system $H_3^{(2+)}$, the longitudinal vibrational
modes of the nuclei are considered, while transversal modes (deviation
from linearity of the configuration as well as alignment along
magnetic line) are not taken into account.  The position of the lowest
longitudinal vibrational state is estimated.

Our calculations demonstrate that for all studied one-electron systems
the binding energy $E_b=\gamma-E_{total}$ (affinity to keep the
electron bound) grows with the increase of the magnetic field strength
for $B=0 - 4.414\times 10^{13} \,G$. For almost all accessible magnetic
fields $B\lesssim 10^{13} \,G$ the molecular ion $H_2^+$ (see the
results in \cite{Lopez:1997}) is the most bound one-electron system
while for magnetic fields $B > 10^{13}\,G$ the molecular ion
$H_3^{(2+)}$ becomes the most bound. Our calculations show as well
that for magnetic fields $B>10^{13} \,G$ the potential energy surface of
the system $(ppppe)$ exhibits the appearance of a minimum indicating
the possible formation of such a molecular ion $H_4^{(3+)}$. It hints
at a certain hierarchy of the thresholds for the appearance of new
one-electron linear molecular system: $H_3^{(2+)}$ -- at $B >
10^{11}G$, $H_4^{(3+)}$ -- at $B > 10^{13}\,G$ etc \footnote{In
  framework of non-relativistic consideration it can be easily checked
  that there exist magnetic fields for which a potential energy
  surface for $H_5^{(4+)}$ etc. can develop a minimum.}.  If this is
so, it is very unlikely that the system $H_5^{(4+)}$ can appear for
magnetic fields existing on the surface of neutron stars. For larger
magnetic fields a reliable study requires taking into account
relativistic corrections, which is beyond of the scope of this
article. So, the question about the possible existence of $H_5^{(4+)}$
is open.

\section{Method}
The present calculations are carried out in the framework of a
variational method using, for each system, a {\it unique} simple
trial function equally applicable to any value of the magnetic
field strength.  For a successful study a wise choice of trial
functions is needed.  A constructive criterion for an adequate
choice of trial function (see \cite{Turbiner:1980}---\cite{Turbiner:1989}) 
is the following: (i) the
trial function $\Psi_t(x)$ should include all symmetry properties
of the problem in hand; (ii) if the ground state is studied, the
trial function should not vanish inside the domain where the
problem is defined; (iii) the potential
$V_t(x)=\frac{{\mathbf\nabla}^2
\Psi_t}{\Psi_t}$, for which the trial function is an exact
eigenfunction, should reproduce the original potential behavior
near singularities as well as its asymptotic behavior.

\begin{figure}[thb]
\caption{One-electron systems in a magnetic field $B$ -- explanation of
  the notations  used. (a) $H$-atom, (b) $H_2^+$, (c)
  $H_3^{(2+)}$, (d) $H_4^{(3+)}$. The positions of charged centers are
  marked by bullets.}
\label{fig:1}
\[
\begin{array}{c}
     {\begin{picture}(155,100)
      \put(0,0){\psfig{figure=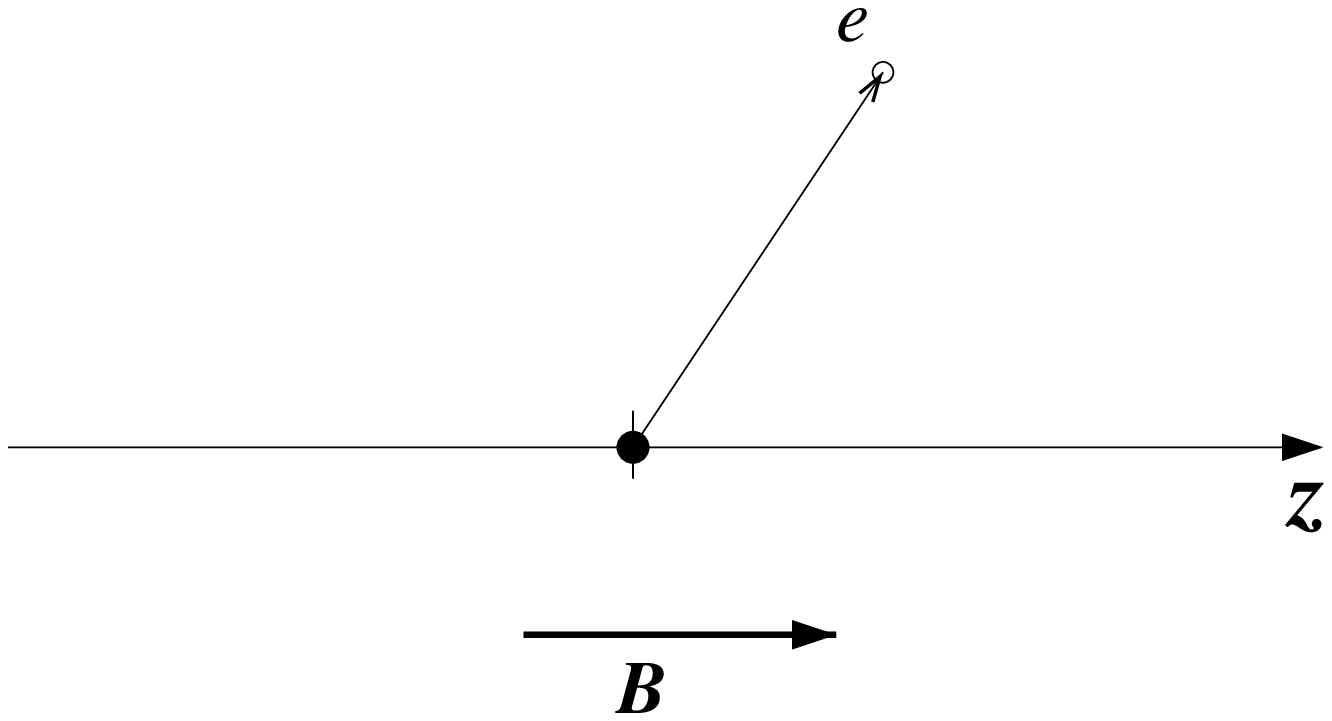,width=2.25in,height=1.2in,angle=-90}}
      \put(100,45){\mbox{$r$}}
      \end{picture}}
\\[5pt]   (a)  \\[5pt]
     {\begin{picture}(155,100)
      \put(0,0){\psfig{figure=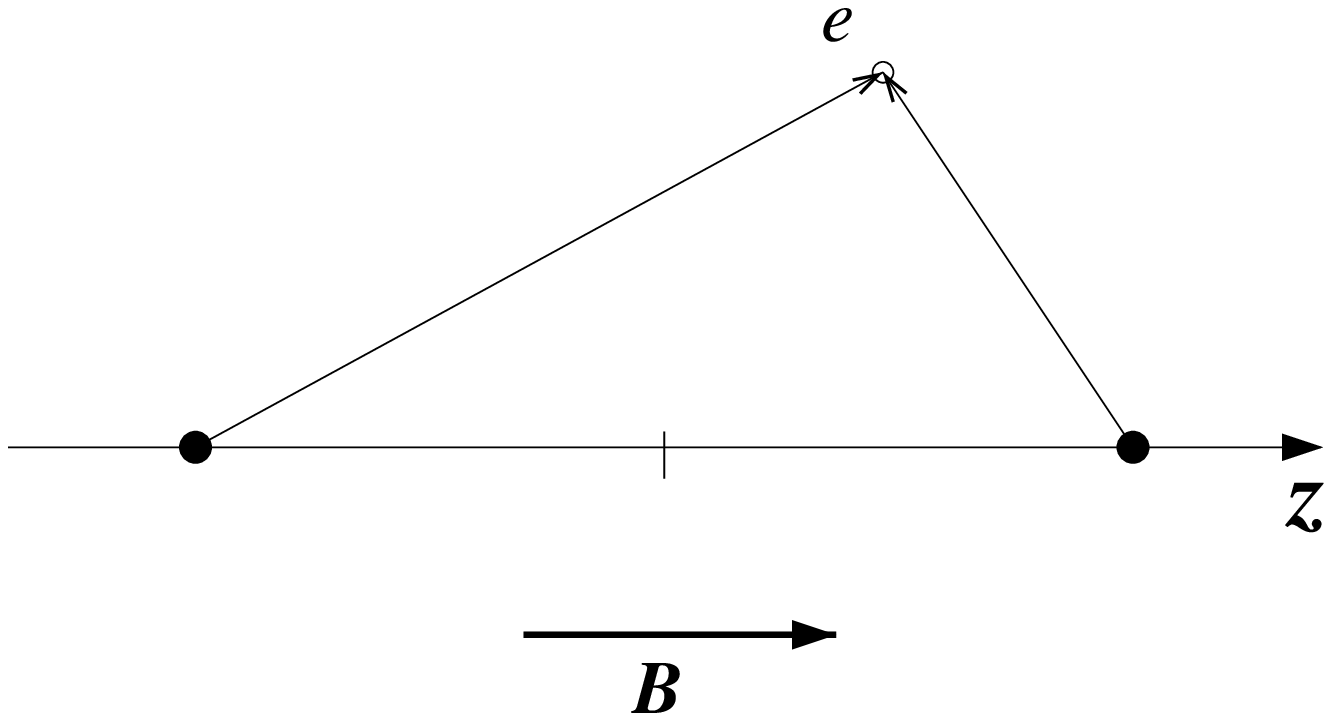,width=2.25in,height=1.2in,angle=-90}}
      \put(50,60){\mbox{$r_1$}}
      \put(130,60){\mbox{$r_2$}}
      \put(7,14){\mbox{$-R/2$}}
      \put(79,14){\mbox{$0$}}
      \put(120,14){\mbox{$+R/2$}}
      \end{picture}}
\\[5pt]  (b)  \\[5pt]
     {\begin{picture}(155,100)
     \put(0,0){\psfig{figure=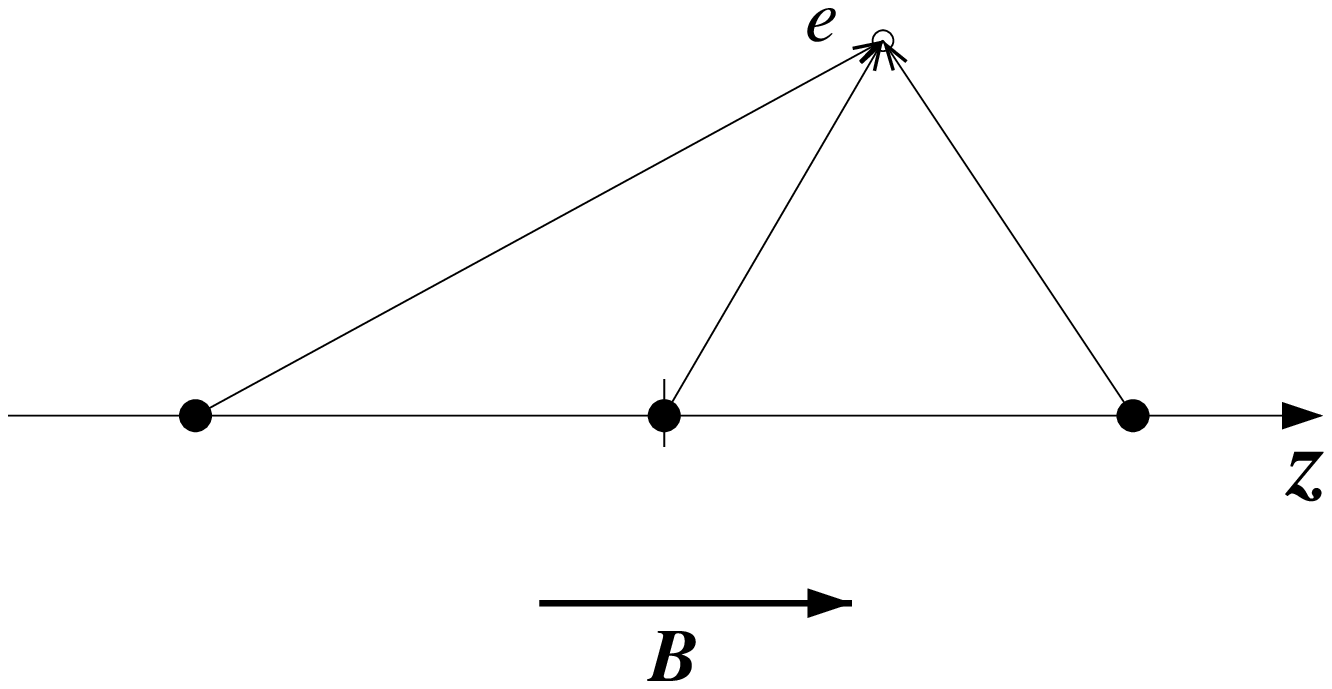,width=2.25in,height=1.2in,angle=-90}}
      \put(50,60){\mbox{$r_1$}}
      \put(100,55){\mbox{$r_2$}}
      \put(130,60){\mbox{$r_3$}}
      \put(7,14){\mbox{$-R_1$}}
      \put(79,14){\mbox{$0$}}
      \put(130,14){\mbox{$R_2$}}
      \end{picture}}
\\[5pt]  (c)  \\[5pt]
      {\begin{picture}(155,100)
      \put(0,0){\psfig{figure=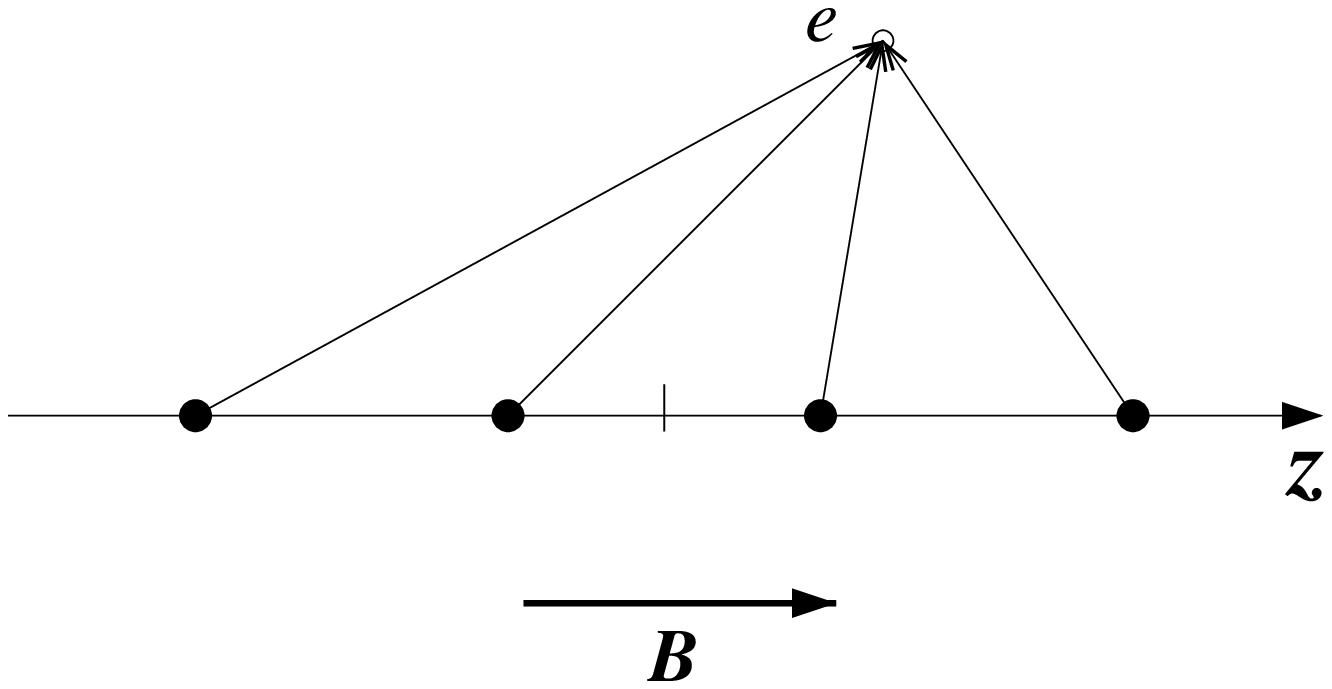,width=2.25in,height=1.2in,angle=-90}}
      \put(50,60){ \mbox{$r_1$}}
      \put(80,45){ \mbox{$r_2$}}
      \put(107,50){\mbox{$r_3$}}
      \put(130,60){\mbox{$r_4$}}
      \put(7,14){  \mbox{$-R_1$}}
      \put(45,14){ \mbox{$-R_2$}}
      \put(92,14){ \mbox{$R_3$}}
      \put(135,14){\mbox{$R_4$}}
      \end{picture}}
\\[5pt]  (d)
\end{array}
\]
\end{figure}

Let us first introduce notation (see Fig. ~\ref{fig:1}).  We consider
identical, infinitely-heavy centers of unit charge situated on the
$z$-axis.  The magnetic field of strength $B$ is directed along the
$z$ axis, ${\vec B} = (0,0,B)$.  Throughout the paper the
Rydberg is used as the energy unit. For the other quantities standard
atomic units are used.

The potentials corresponding to the systems we study are given by

\begin{equation}
\label{eq:1}
\begin{array}{lcl}
\displaystyle
V = -\frac{2}{r}  + \frac{B^2 \rho^2}{4} \ , && \mbox{$H$} \\[10pt]
\displaystyle
V = \frac{2}{\st R} -\frac{2}{r_1} -\frac{2}{r_2}  +
\frac{B^2 \rho^2}{4} \ , && \mbox{$H_2^+$}  \\[10pt]
\displaystyle
V=\frac{2}{\st R_1} + \frac{2}{\st R_2} + \frac{2}{\st R_1+R_2}
-\frac{2}{r_1} - \frac{2}{r_2} - \frac{2}{r_3} +
\frac{B^2 \rho^2}{4}  \ , &&  \mbox{$H_3^{(2+)}$}   \\[10pt]
\displaystyle
V=\frac{2}{\st R_1-R_2} + \frac{2}{\st R_1+R_3} + \frac{2}{\st R_1+R_4}
+\frac{2}{\st R_2-R_3} + \frac{2}{\st R_2+R_4} + \frac{2}{\st R_4-R_3}
-\frac{2}{r_1} - \frac{2}{r_2} - \frac{2}{r_3} - \frac{2}{r_4} +
\frac{B^2 \rho^2}{4} \ ,  &&  \mbox{$H_4^{(3+)}$}
\end{array}
\end{equation}
where the quantity $\rho = \sqrt{x^2+y^2}$ is the distance from the
electron to the $z$-axis, and the $r_{i}$ are the distances from the
electron to the {\it i}th center.

\section{Trial Wave Functions}
A class of trial functions which satisfy the above-mentioned
criterion incorporating the basic physical features of Coulomb
systems in a magnetic field directed along the $z$-axis contains a
function 
\begin{eqnarray}
  \label{eq:2}
  \psi_0 &=& {\rm e}^{-\varphi(r_1,\ldots r_N, \rho)}
  ,\nonumber \\
\varphi(r_1,\ldots r_N, \rho) &=&
\sum_{i=1}^{N} \alpha_i\, r_i + \beta \, B \rho^2/4
\end{eqnarray}
as a basic element,  
where $\alpha_i$ and $ \beta$ are variational parameters, and $N$
is the number of centers. This function is nothing but the
product of ground-state Coulomb functions and the lowest Landau
orbital and  is the exact eigenfunction
of the potential 
\begin{eqnarray}
  \label{eq:3}
V_0 &=&
  \frac{\Delta \psi_0}{\psi_0} =
  \left[- \nabla^2 \varphi +  (\nabla \varphi )^2  \right]
  \nonumber \\
&=&
-2\sum_i \alpha_i \frac{1}{r_i} - \beta B  +
\sum_{i,j} \alpha_i \alpha_j \, ( {\hat n}_i \cdot {\hat n}_j )
+
\sum_i \alpha_i \beta B \, \frac{ \rho^2}{r_i}
+ {\beta}^2 \frac{B^2\rho^2}{4}\ ,
\\[5pt]
{\hat n}_i \cdot {\hat n}_j &=& \frac{1}{r_i r_j}
\left[
\rho^2 
+ (z-z_i)(z-z_j)
\right] \ ,
\nonumber
\end{eqnarray}
where ${\hat n}_i$ is unit vector in $i$th direction. This potential
reproduces  Coulomb-like behavior near the centers and 
two-dimensional oscillator behavior in the $(x,y)$ plane at large
distances (cf. (1) ).

Since the centers are identical the problem possesses  permutation
symmetry. Hence the ground state trial function must be
permutationally-symmetric with respect to permutation of positions
of the centers, $r_i \leftrightarrow r_j$.

One of the simplest functions satisfying the above-mentioned
criterion is the function (\ref{eq:2}) itself  with the same $\al$'s
(those can be placed equal to $\al_1$):
 \begin{equation}
 \label{eq:4}
 \Psi_1= {e}^{-\al_1  \left({ \sum_{i=1}^{N}} \,r_i\right)
  - \be_1 B \rho^2 /4},
 \end{equation}
where $\al_1,\be_1$ are variational parameters. Actually, this 
function is a Heitler-London type function multiplied by the
lowest Landau orbital. If the internuclear distances are taken as
parameters, it has in total ($N+1$) variational parameters. It is
quite natural from a physical viewpoint to assume that this
function gives a description of the system near the equilibrium
position. The potential
$V_1(x)=\frac{{\mathbf\nabla}^2\Psi_1}{\Psi_1}$, corresponding to
this function is:
\begin{equation}\label{eq:5}
V_1 \ =\ N \al^2_1 -B \be_1 -2 \al_1 \,\sum_{i=1}^{N}
\Bigg(\frac{1}{r_i}\Bigg)+\frac{\be^2_1
B^2\rho^2}{4}+
2\al^2_1 \sum_{i<j} \, {\hat n}_i \cdot {\hat n}_j
+\ \al_1  \be_1 B\rho^2
\,\sum_{i=1}^{N}
\Bigg(\frac{1}{r_i}\Bigg) \ ,
\end{equation}
(cf. (3)).
 
Another possible trial function is a permutationally-symmetric linear
superposition of the functions (\ref{eq:2}) in which all $\al$'s are
placed to be equal to zero except one, which is denoted $\al_2$
\begin{equation}
\label{eq:6}
\Psi_2=
\Big( \sum_{i=1}^{N} {e}^{-\al_2  r_i} \Big)
{e}^{-\be_2 B \rho^2 /4} \ .
\end{equation}
Here $\al_2,\be_2$ are variational parameters. This function is a
Hund-Mulliken type function multiplied by the lowest Landau orbital.
One can assume that when all internuclear distances are sufficiently
large this function dominates, giving an essential contribution. Thus,
it describes  an interaction of the hydrogen atom with $(N-1)$ charged
centers. This function can also describe a possible decay mode of the
studied system to the hydrogen atom and $(N-1)$ protons. As in
Eq.(\ref{eq:4}), the trial function (\ref{eq:6}) is characterized by
$(N+1)$ variational parameters.

Another trial function describes a possible decay mode
$H_N^{((N-1)+)} \rar H_2^{+} + (N-2)p,\ N \geq 3$, and is a 
permutationally-symmetric linear superposition of the functions
(\ref{eq:2}) with all $\al$'s are equal to zero except two, which
are taken to be equal and  denoted as $\al_3$:
\begin{equation}
\label{eq:7}
\Psi_3=
\Big( \sum_{i<j} {e}^{-\al_3  (r_i+r_j)}\Big)
{e}^{-\be_3 B \rho^2 /4} \ ,
\end{equation}
where $\al_3,\be_3$ are variational parameters. It seems that the
function (\ref{eq:7}) gives the dominant contribution to the large
internuclear distances when the distance between two neighboring
centers is kept fixed. It can be true for magnetic fields $B <
10^{13}\,G$, where the system with the lowest total energy is $H_2^+$.
Eq.(\ref{eq:7}) also depends on $(N+1)$ variational parameters.

Finally, for magnetic fields $B>10^{11}\,G$ where the exotic state
$H_3^{(2+)}$ begins to appear we also introduce a trial function
of the form
\begin{equation}
\label{eq:8}
\Psi_4=
\Big( \sum_{i<j<k} {e}^{-\al_4  (r_i+r_j+r_k)}\Big)
{e}^{-\be_4 B \rho^2 /4} \ .
\end{equation}
where again $\al_4$ and $\be_4$ are variational parameters. This
function depends on $N+1$ variational parameters.

In order to incorporate in a single trial function the behavior of the
system both near equilibrium and at large distances, we use an
appropriate interpolation of the trial functions (\ref{eq:4}),
(\ref{eq:6}), (\ref{eq:7}) and (\ref{eq:8}). There are several
natural types of such interpolations:

\begin{itemize}
\item[(i)] Total non-linear superposition of the form:
\begin{equation}
    \label{eq:9}
    \Psi_{5-nls}=
    \left(
      \sum_{\{ \al_1^\prime, \ldots \al_N^\prime\} }
     \hspace{-15pt}
     {e}^{ -\al_1^\prime\, r_1 - \al_2^\prime\, r_2 - \ldots
                              - \al_N^\prime\, r_{N}  }
    \right) {e}^{-\be^\prime  B {\rho^2}/{4}    }
\end{equation}
where $\al_1^\prime, \ldots \al_N^\prime $ and $\be^\prime$ are
variational parameters, and the sum is over all permutations of
the parameters $\{ \al_1^\prime, \ldots \al_N^\prime\}$. The
function (\ref{eq:9}) is an $N$-center modification of the
Guillemin-Zener type function used for the description of the
molecular ion $H^+_2$ in a magnetic field (see Eq.(2.5) in
\cite{Lopez:1997}).  If all parameters coincide $\al_{i}^\prime =
\al_{1}, \ i=1 \ldots N$, the function (\ref{eq:9}) reduces to
the Heitler-London type  function (\ref{eq:4}). If the only
parameter is non-zero, say $\al_1^\prime =\al_2$, this 
function reduces to the Hund-Mulliken type wave function
(\ref{eq:6}), if two parameters are non-zero, and equal, say,
$\al_1^\prime = \al_2^\prime = \al_3$ it reduces to the trial 
function (\ref{eq:7}), and if three parameters are non-zero,
$\al_1^\prime = \al_2^\prime =\al_3^\prime = \al_4$ it reduces to
(\ref{eq:8}).  In general, there are $2N$ variational parameters
(including the internuclear distances as parameters)
characterizing the trial function (\ref{eq:9}).

\item[(ii)] Partial non-linear superpositions:\\
these are special cases of (\ref{eq:9}) when  two or more
parameters are equal. For instance, for $N=3$
 $\al_{1}^\prime=\al_{2}^\prime \ne \al_{3}^\prime $ :
\begin{equation}
\label{eq:10}
\Psi_{6-{nls-p}}= \Big(
 {e}^{-\al_1^\prime  (r_1 + r_2) - \al_3^\prime  r_3} +
 {e}^{-\al_1^\prime  (r_1 + r_3) - \al_3^\prime  r_2 } +
 {e}^{-\al_1^\prime  (r_2 + r_3) - \al_3^\prime  r_1 }
 \Big) {e}^{-\be^\prime B\rho^2/4}\ ,
\end{equation}
This function can be considered as a non-linear interpolation
between Eqs. (\ref{eq:6}) and (\ref{eq:7}) for the $H_3^{(2+)}$
ion.

\item[(iii)] Linear superposition of Eqs.(\ref{eq:4}), (\ref{eq:6}),
(\ref{eq:7}), (\ref{eq:8})
\begin{equation}
\label{eq:11}
\Psi_{7-{ls}}= A_1 \Psi_1 + A_2 \Psi_2 + A_3 \Psi_3 + A_4 \Psi_4\ ,
\end{equation}
where $A_1, A_2, A_3, A_4$  are taken as extra variational
parameters. It should be mentioned that in general
$\Psi_{1,2,3,4}$ are not orthogonal and thus the parameters
$A_{1,2,3,4}$ do not have a meaning of the weight factors.

\end{itemize}

So, each of the above interpolations has a certain physical meaning
describing different physical situations.  In order to make up a rich
description of the systems one can take a linear superposition of all
previously mentioned functions with factors in front of them, those
are treated as extra variational parameters.  In this work we use
only some particular cases of this general interpolation procedure
(see below for details).

The minimization procedure is carried out using the standard
minimization package MINUIT from CERN-LIB. We use the
integration routine D01FCF from NAG-LIB. All integrals are
calculated with relative accuracy $\gtrsim 10^{-8}$. 

\section{Results}

The results are summarized in Table I. In general, as a common feature
the energy dependences for the hydrogen atom $H$ and molecular ions
$H_2^+$ and $H_3^{(2+)}$ are characterized by the growth of the
binding energy, as well as a decrease of the equilibrium distances
($R$ for $H_2^+$, and $R_1 + R_2$ for $H_3^{(2+)}$) with increase of
magnetic field strengths \footnote{It is known that in the limit of
  large magnetic fields the binding energy of electron in $H$,
  $H_2^+$, $H_n$ systems as well as in heavy atoms behaves $\propto
  (\log \gamma)^2 $ and the equilibrium distances $\propto (\log
  \gamma)^{-1}$ (see, for instance,
  \cite{Kadomtsev:1971,Ruderman:1971,Salpeter:1992,Garstang:1977} and
  discussion therein). Similar estimates should hold for other
  one-electron linear systems.}. It seems reasonable to conjecture
that this behavior persists for magnetic fields $B > 4.414 \times
10^{13}\,G$ for which a non-relativistic treatment is no longer
applicable.  In Table I \(L\) is the ``natural'' size of the system
which is determined by the internuclear distances: $L=R$ for $H_2^+$,
$L=R_1+R_2$ for $H_3^{(2+)}$ and $L=R_1+R_4$ for $H_4^{(3+)}$, see
Fig.  1.  For $H_3^{(2+)}$, the energy of the lowest vibrational state
is defined as $\delta E = \frac{1}{2} \hbar\omega_+ + \frac{1}{2}
\hbar\omega_-$, where $\omega_\pm$ are determined by curvatures near
minimum of the potential well along the principal axes: $ {\cal R}_+ =
R_1 + R_2, \ {\cal R}_- =R_2 - R_1$, see Fig. 1. One should emphasize
that with a magnetic field growth all studied systems become more and
more bounded. Also for both $H_2^{+}$ and $H_3^{(2+)}$ the energy
of the lowest vibrational state increases. The vibrational state for
$H_4^{(3+)}$ was not studied.

\subsection{$H$-Atom}
Recently, an extensive numerical analysis of different features of the
hydrogen atom in constant magnetic field was presented in the book by
Ruder et al. \cite{Ruder}. In our study 
for the ground state of the $H$-atom we use the trial function of the form
(\ref{eq:2}) at $N=1$ (see \cite{Turbiner:1984}---\cite{Turbiner:1989})
\[
\psi_H = {\rm e}^{-\al r -\be B\rho^2/4} ,
\]
where $\al$ and $\be$ are variational parameters. A comparison with
the results obtained by Lai et al. \cite{Salpeter:1992} shows that this
simple trial function gives  a relative accuracy  $10^{-4}$ in energy 
for magnetic fields $B=0 - 4.414\times 10^{13} \,G$. The
binding energy for this system increases with magnetic field growth
and the longitudinal size reduces drastically from $1.5\ a.u.$ at
$B=0$ to $0.242\ a.u. $ at $B=4.414\times 10^{13} \,G$. The total energy
for the $H$-atom is larger than the total energy for $H_2^+$ for all
studied magnetic fields, as well as larger than the total energies for
$H_3^{(2+)}$ and for $H_4^{(3+)}$ for $B\gtrsim 10^{11} \,G$  and $B
\gtrsim 4.414\times 10^{13} \,G$ , respectively (see Table I). So, the
$H$-atom is a least bound one-electron  system among $H_2^+$,
$H_3^{(2+)}$ and $H_4^{(3+)}$.

\subsection{$H_2^+$ molecular ion}

For this system a linear superposition of trial functions
(\ref{eq:4}), (\ref{eq:6}) and (\ref{eq:9}) for $N=2$ is used for
present variational calculation (cf.  \cite{Lopez:1997}).  In total
there are ten variational parameters and the present calculation provides
the most accurate ground state total energy for $B\gtrsim 10^9 \,G$. The
ground state corresponds to the positive parity state $1\sigma_u$. For
magnetic fields in the region $B=0-10^{13}\,G $ this system is
characterized by the lowest total energy among the one-electron
systems we consider, being the most stable.  The binding energy
increases (and the equilibrium distance $R$ decreases) with the
increase of the magnetic field. It is worth noting that the
longitudinal localization length of the electron $\langle | z |
\rangle$ is larger than the ``natural'' size of the system $L=R$ for
magnetic fields $B>0$ (See Figs. 2,3). The energy of the lowest longitudinal
vibrational level is estimated using a harmonic approximation of the
bottom of the potential energy curve. The position of the lowest
vibrational state in comparison with the Born-Oppenheimer minimum of
the total energy grows steadily from $\delta E = 0.01 \, Ry $ at $B=0$ to
$\delta E = 1.23 \, Ry$ for $B=4.414\times 10^{13} \,G $, respectively.

\subsection{$H_3^{(2+)}$ molecular ion}

For this exotic system $H_3^{(2+)}$ theoretical evidence was provided
of its existence for magnetic fields $B\gtrsim 10^{11} \,G$
\cite{Turbiner:1999}. For the present variational calculations we use
a linear superposition of trial functions (\ref{eq:4}), (\ref{eq:6}),
(\ref{eq:7}) and (\ref{eq:9}) for $N=3$.  We have in total fourteen
variational parameters. As in the case of $H_2^{+}$ the binding energy
increases, and the natural size of the system $L=R_1 + R_2$ decreases,
as the magnetic field grows. In contrast to the $H_2^+$ ion, the
longitudinal localization length of the electron $\langle | z |
\rangle$ is smaller than the natural size of the system. In this case
the electronic cloud mainly surrounds the central proton (See Figs.
2,3).  The potential energy surface on the plane $(R_1,R_2)$ exhibits
a well-pronounced minimum at $R_1=R_2$ starting from $B \sim 10^{11}
\,G$.  The energy of the lowest longitudinal vibrational state is
calculated using a harmonic approximation around the minimum along the
principal axes ${\cal R}_+ = R_1 + R_2$, ${\cal R}_- = R_2 - R_1$ in
the plane $(R_1,R_2)$. Our results show that for $B=10^{11} \,G$ the
energy of the lowest longitudinal vibrational state in harmonic
approximation $\delta E = 0.189 \, Ry$ lies much above the top of the
barrier $\Delta E = 0.054 \, Ry$.  However, the situation changes with
growth of magnetic field. For $B \gtrsim 10^{12} \,G$ the potential well
becomes sufficiently deep to keep the lowest vibrational state inside
of the well (see Table I). For $B=10^{11} - 10^{13}\, G$ the system
$H_3^{(2+)}$ is unstable towards a decay to $H_2^{+} + p $. With the
growth of magnetic field the total energy of $H_3^{(2+)}$ becomes
lower than the total energy for $H_2^+$. This happens at $B \approx
4.414\times 10^{13} \,G$ and it means that for $B \gtrsim 4.414\times
10^{13} \,G$ the exotic molecular ion $H_3^{(2+)}$ becomes the most
bound one-electron system.

\subsection{$H_4^{(3+)}$ molecular ion}

To study this system we use a superposition of the trial functions
(\ref{eq:4}), (\ref{eq:6}), (\ref{eq:7}) and (\ref{eq:8}) for $N=4$ in
total with fifteen variational parameters. A symmetry argument implies
that the minimum, if it exists, should correspond to the configuration
$R_3=R_2$ and $R_4=R_1$ (see Fig.~\ref{fig:1}). The results show that
for magnetic fields $B \lesssim 10^{13} \,G$  the potential energy
surface in the plane $(R_1,R_2)$ does not exhibit any minimum. The first
indication to a possible existence of the minimum comes at $B =
10^{13}\,G$ and for $B=4.414\times 10^{13} \,G$  the potential energy
surface indicates the appearance of a well-pronounced minimum of 
depth $\Delta E = 0.4 \, Ry$ at $R_1=0.225 $ and $R_2=0.058$. The
longitudinal localization length of the electron $\langle | z |
\rangle$ is smaller than the natural size of the system $L= 2R_1$
similarly to what takes place for $H_3^{(2+)}$.  The electronic cloud
surrounds mainly the two central protons (see Figs. 2,3) similar to
$H_3^{(2+)}$. It is important to stress
that the total energy for this system is smaller than the total energy
of the $H$ atom but is larger than the total energy for $H_2^+$ and
$H_3^{(2+)}$ at $B=4.414\times 10^{13} \,G$.

\section{Conclusions}

In the presence of a strong magnetic field the formation of exotic
chemical compounds can occur. In particular, in present article we
demonstrate that in addition to the $H$-atom and the $H_2^+$ molecular
ion two more one-electron systems $H_3^{(2+)}$ and $H_4^{(3+)}$ can
exist which do not exist without a magnetic field.  The results of our
consideration show an appearance of a minimum in the potential energy
surfaces for the systems $(pppe)$ and $(ppppe)$ for magnetic fields
$B>10^{11} \,G$ and at $B=4.414\times 10^{13} \,G$, respectively,
indicating the formation of the exotic ions $H_3^{(2+)}$ and
$H_4^{(3+)}$.  Among these systems the molecular ion $H_2^+$ possesses
the lowest total energy in the region $B \lesssim 10^{13} \,G$ being the
most stable towards a fission of proton(s) or an electron. The ion
$H_3^{(2+)}$ can decay to $H_2^{+} + p$ but not to $H + p + p$.
However, for magnetic fields $B \gtrsim 4.414\times 10^{13} \,G$ the ion
$H_3^{(2+)}$ becomes the most stable one-electron system possessing
the lowest total energy.  Our estimates show that the lowest
longitudinal vibrational state of $H_3^{(2+)}$ can also exist for
magnetic fields $B\ge 10^{12} \,G$.

For a magnetic field $B\simeq 4.414\times 10^{13} \,G$, where we assume
that a non-relativistic treatment is still valid, the exotic molecular
ion $H_4^{(3+)}$ can exist. This ion is unstable towards a decay to
$H_2^+ + p + p$ and $H_3^{(2+)} + p$, however having no decay mode 
$H + p + p + p$. A further study of this system beyond the
non-relativistic treatment is needed for reliable evidence of its
existence. We should also mention that there is a general question
about applicability of the Born-Oppenheimer
approximation for the  systems in a magnetic field (see discussion in
\cite{Schmelcher,Ruder,Salpeter:1995}). In all our calculations we
use the standard Born-Oppenheimer approximation.   

\acknowledgments

The authors wish to thank K.G.~Boreskov for useful
discussions and M. Ryan for reading the manuscript. 

This work is supported in part by DGAPA Grant 
\# IN105296 (M\'exico).


\newpage

\newpage

\begin{figure}[tb]
\caption{Electronic probability density for  (a) $H_2^+$, (b)
  $H_3^{(2+)}$ and  (c) $H_4^{(3+)}$ for $B=4.414\times 10^{13}
  \,G$. Normalization of $|\Psi|^2$ is not fixed.}
\label{Fig:2}
\[
\begin{array}{c}
\psfig{figure=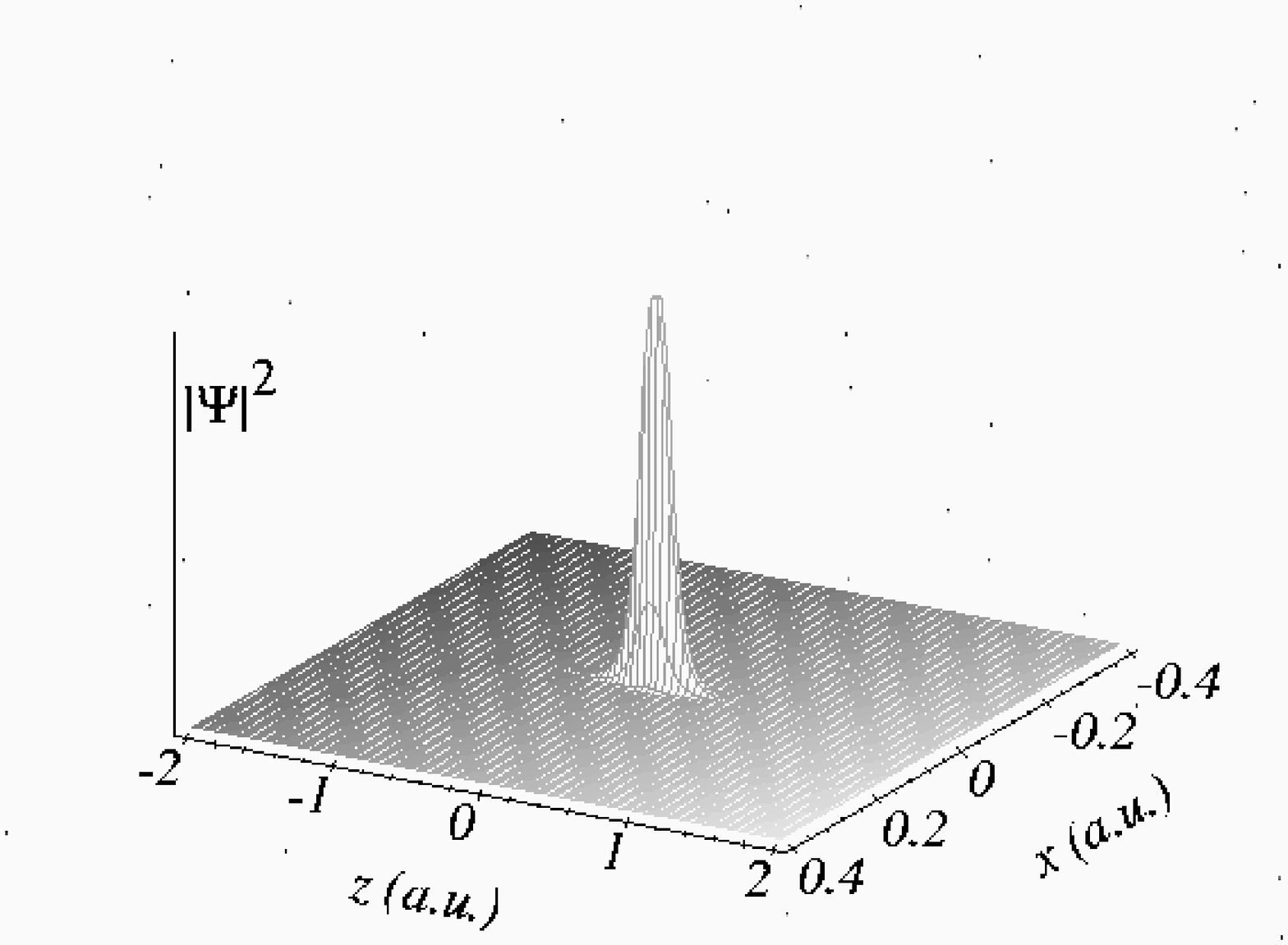,width=3.2in,angle=-90} \\
(a)\\
\psfig{figure=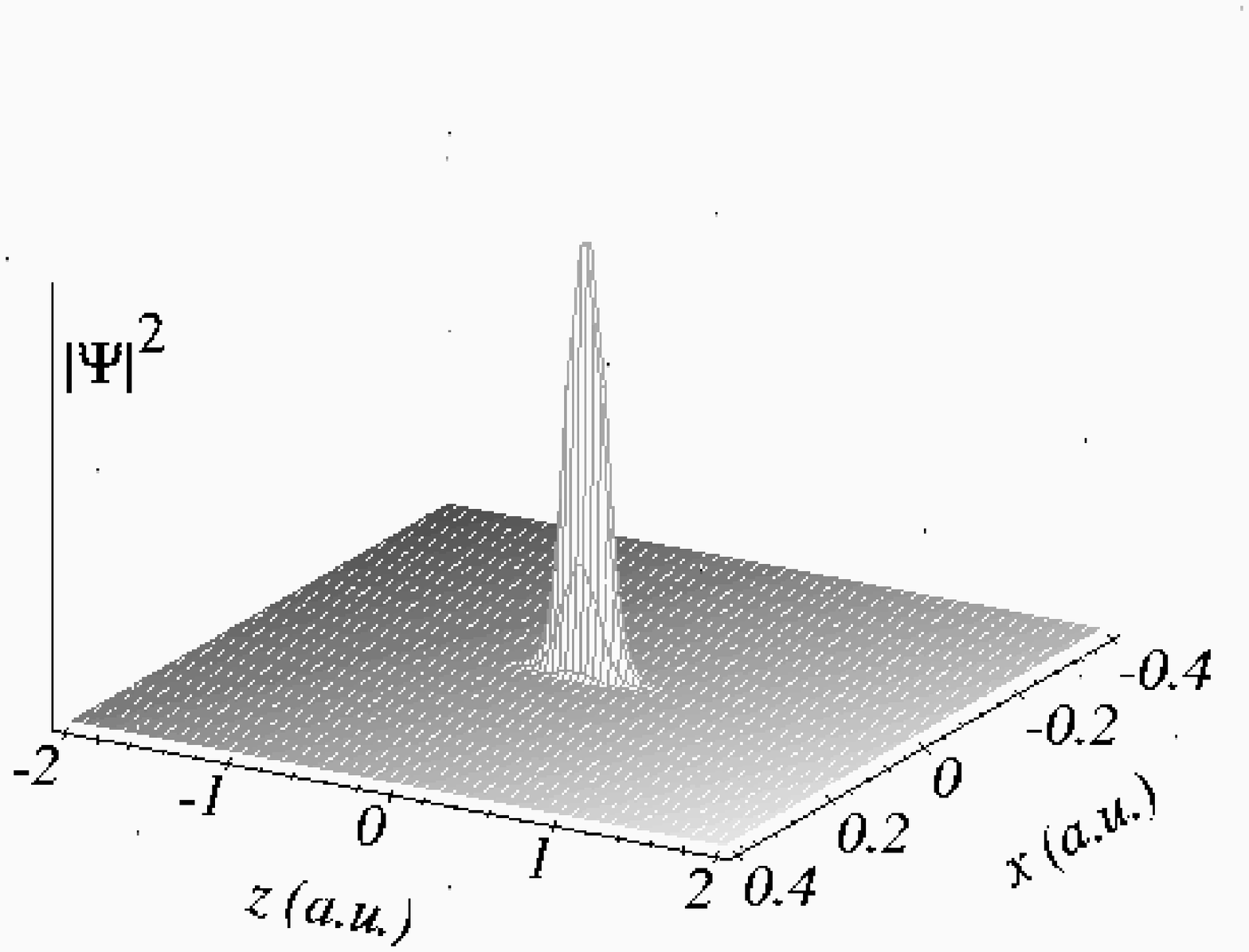,width=3.2in,angle=-90} \\
(b)\\
\psfig{figure=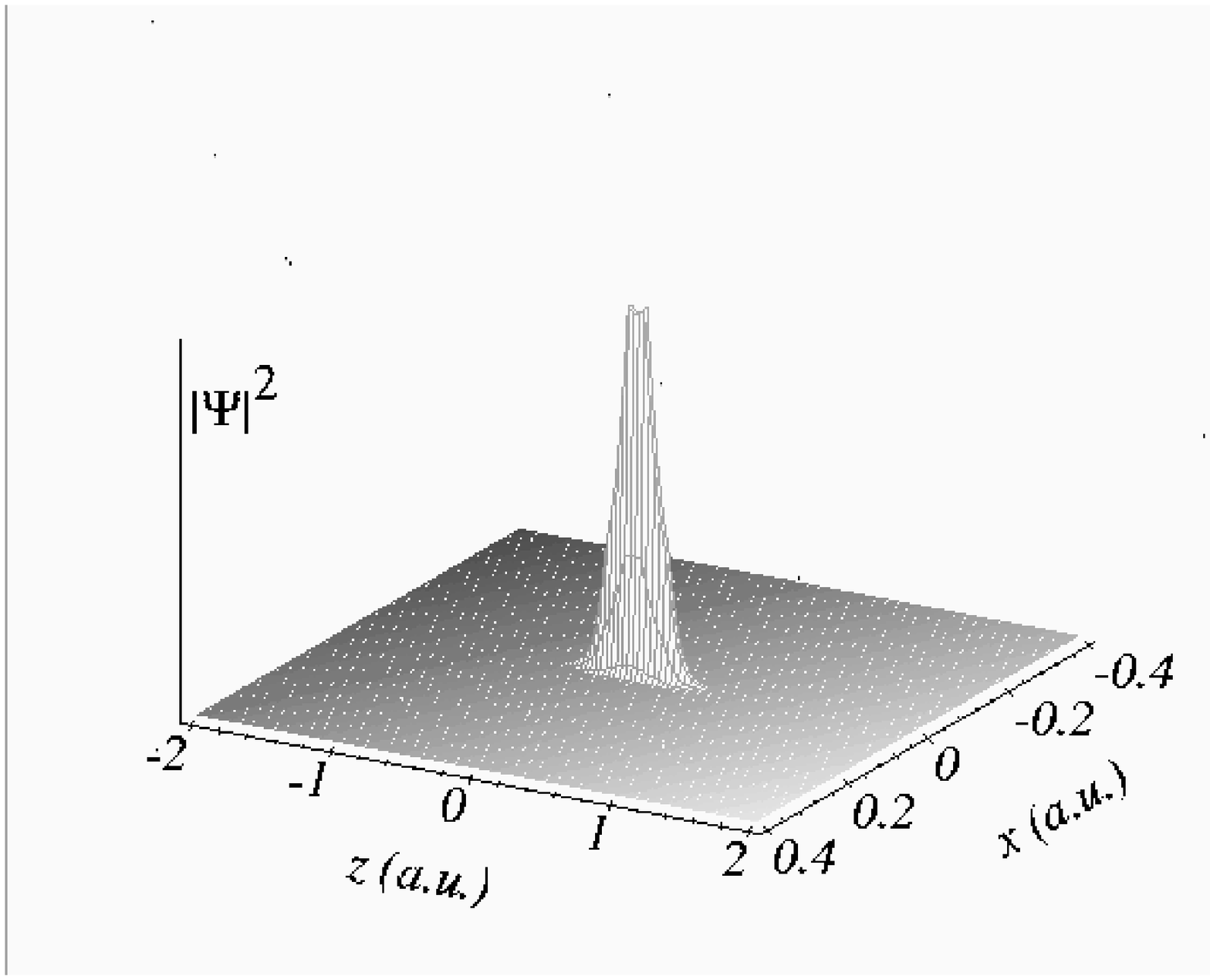,width=3.2in,angle=-90} \\
(c)
\end{array}
\]
\end{figure}

\newpage

\begin{figure}[tb]
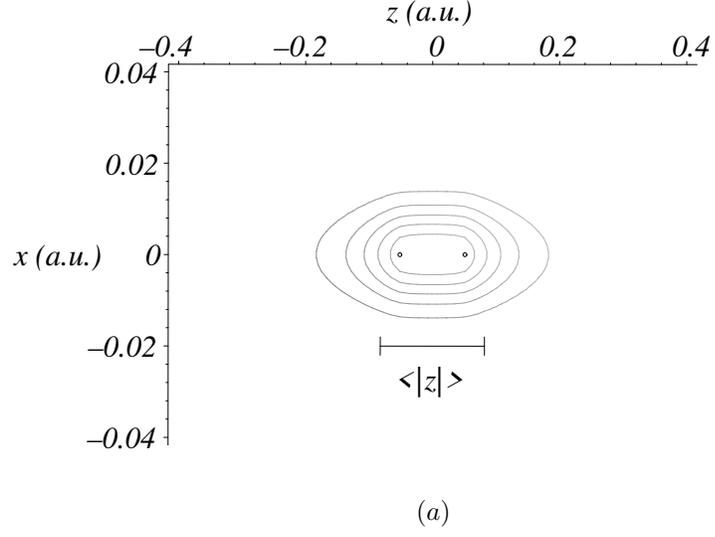

\caption{Electronic probability density contours for  (a) $H_2^+$, (b)
  $H_3^{(2+)}$ and  (c) $H_4^{(3+)}$ for $B=4.414\times 10^{13}\,G$
 . The position of the centers is indicated by small circles. The
  longitudinal localization length of the electron $\langle | z | \rangle$ is
  displayed.}
\label{fig:3}
\[
\begin{array}{c}
\hspace{-70pt}\psfig{figure=psi2_h2+c.ps,width=2.5in,angle=-90}  \\[-70pt]
(a)\\[25pt]
\hspace{-70pt}\psfig{figure=psi2_h3++c.ps,width=2.5in,angle=-90} \\[-70pt]
(b)\\[25pt]
\hspace{-70pt}\psfig{figure=psi2_h4+++c.ps,width=2.5in,angle=-90}\\[-70pt]
(c)
\end{array}
\]
\end{figure}

\pagebreak

\newpage

\begin{table}
\label{table:1}
\caption{Variational results for the one--electron linear systems
$H$, $H_2^+$, $H_3^{(2+)}$ and $H_4^{(3+)}$ 
in a strong magnetic 
field for total energy $E$, the natural size of the system $L$ (see text),
longitudinal electron localization length $\langle | z | \rangle$, 
longitudinal vibrational energy $\delta E$ measured from
the bottom of the well, and the height of the barrier $\Delta E$ (see
text for further  explanations). Results in parenthesis are taken 
from  ${}^\star$ Lai et al.
\protect\cite{Salpeter:1992}, ${}^{\star\star}$ Lai et al. 
\protect\cite{Salpeter:1995}, ${}^{\dag}$ Wind
\protect\cite{Wind:1965}, ${}^\ddagger$ Wille
\protect\cite{Wille:1988},  ${}^{\dag\dag}$ Kravchenko and Liberman
\protect\cite{Kravchenko:1997}, ${}^{\diamond}$
\protect\cite{Lopez:1997}, ${}^{\diamond\diamond}$
\protect\cite{Turbiner:1999}
.} 
 \begin{tabular}{cccccc} 
      \( B \) (Gauss) & &
      \( H \)& \( H_{2}^{+} \) & \( H_{3}^{(2+)} \)& \( H_{4}^{(3+)} \) \\
    & & & & &  \\ \hline \hline
  & \( E \) & -1.000 & -1.20525${}^{\diamond}$               & --- &        --- \\
  &         & ---    & (-1.205268${}^{\dag}$) & --- &        --- \\
  & \( L \) & ---    &  1.997                 & --- &        --- \\
  &         & ---    & (2.000${}^{\dag}$)     & --- &        --- \\
\( 0 \)
  & \( \langle |z| \rangle \) &$1.50$ &1.704 & --- & --- \\
  & \( \delta E \)& --- & $0.0107$              & --- &   --- \\
  &               & --- & ($0.0106\,{}^{\dag\dag}$) & --- &   --- \\
  & \( \Delta E \)& --- & ---      & --- &   ---\\      \hline
  & \( E \)& 36.996             & 35.036${}^{\diamond}$                & 36.429${}^{\diamond\diamond}$  &   ---\\
  &        & (36.929${}^\star$) & (35.043${}^\ddagger$) & --- &        --- \\
  & \( L \)& ---                & 0.593                 & 1.606   &   ---\\
  &        & ---                & (0.593${}^\ddagger$)  & --- &        --- \\
\( 10^{11} \)
& \( \langle |z| \rangle \)     &$0.685$  & 0.624   & 0.864   &  ---\\
& \( \delta E \)                & ---     & $0.084$ & $0.189$ &  ---\\
&                      & ---    & ($0.074\,{}^{\star\star}$)  & --- &        --- \\
& \( \Delta E \)                & ---     & ---     & 0.054   &  ---\\ \hline
& \( E \) &  413.94             &  408.30${}^{\diamond}$               & 410.296${}^{\diamond\diamond}$ &      ---\\
&         &  (413.57${}^\star$) & (408.56${}^\ddagger$) & --- &        --- \\
& \( L \) &    ---              &  0.283                & 0.692   &      ---\\
  &         &  ---              & (0.278${}^\ddagger$)  & --- &        --- \\
\( 10^{12} \)
& \( \langle |z| \rangle \)& $0.441$ & 0.348 &  0.438 &   ---\\
& \( \delta E \)& --- &$0.259$ &$1.033$ &    ---\\
&               & --- &($0.243\,{}^{\star\star}$)  & --- &        --- \\
& \( \Delta E \)& --- & --- & 0.833&          ---\\  \hline
& \( E \)   & 4232.6             & 4218.665${}^{\diamond}$ & 4220.090${}^{\diamond\diamond}$    &          ---\\
&           & (4231.6${}^\star$) & (4218.674${}^\star$)   & --- &  --- \\
& \( L \)   &  ---               & 0.147    & 0.330    &      ---\\
&           &  ---               & (0.15${}^\star$)  & --- &        --- \\
\( 10^{13} \)
& \( \langle |z| \rangle \)&$0.301$ & 0.214& 0.242&   ---\\
& \( \delta E \)& --- &$0.714$ &$2.406$ &              ---\\
&               & --- &($0.625\,{}^{\star\star}$)        & --- &        --- \\
& \( \Delta E \)& --- &---     & 3.58&                     ---\\ \hline
& \( E \)&$18747.4$ & 18724.501& 18723.88&      18735.70  \\
& \( L \)& ---& 0.102& 0.220&   0.450      \\
\( 4.414\times 10^{13} \)
& \( \langle |z| \rangle \)&$0.242$ &$0.164$ & $0.168$& $0.224$ \\
& \( \delta E \)&--- &$1.233$ &$4.27$ &   ---                 \\
& \( \Delta E \)&--- &--- &  $ 6.5 $ &   $0.4$ \\ 
\end{tabular}
\end{table}

\end{document}